\documentclass[%
reprint,
superscriptaddress,
 amsmath,amssymb,
onecolumn
]{revtex4-2}

\usepackage{graphicx}
\usepackage{xcolor}
\usepackage{bm}
\usepackage{multirow}
\definecolor{green}{rgb}{0.3,0.7,0.3}

\newcommand{\ks}{\textcolor{black}}
\newcommand{\kss}{\textcolor{black}}

\begin{document}

\title{Recirculation regions in wakes with base bleed}

\author{K. Steiros}
\email{k.steiros13@imperial.ac.uk}
\affiliation{Department of Aeronautics, Imperial College London, London SW7 2AZ, UK}
\author{N. Bempedelis}
\affiliation{Department of Aeronautics, Imperial College London, London SW7 2AZ, UK}
\affiliation{Department of Mechanical Engineering, University College London, London WC1E 7JE, UK}
\author{L. Ding}
\affiliation{Department of Mechanical and Aerospace Engineering, Princeton University, Princeton, NJ 08544, USA}
\date{\today}

\begin{abstract}
The appearance of detached recirculation regions in wakes with base bleed determines the aerodynamic properties of many natural organisms and technological applications. In this work we introduce an analytical model which captures certain key dimensions of the recirculation region in the wake of porous plates of infinite aspect ratio, along with the porosity range \ks{over which} it exists, when vortex shedding is absent or suppressed. The model is used to interpret why the recirculation region (i) emerges, (ii) migrates away from the body with increasing base bleed, (iii) disappears at a critical bleed and (iv) is partially insensitive to variations in the Reynolds number. The model predictions show considerable agreement with data from laboratory experiments and numerical simulations.
\end{abstract}

\maketitle
\section{Introduction}
It is well known that the near wake of solid bluff bodies is characterized by an attached recirculation region, which determines much of the aerodynamic properties of the configuration \citep{roshko1993perspectives}. An interesting phenomenon occurs when fluid bleeding is superimposed on the wake, i.e. when the body is permeable, or when fluid is actively injected at its base; the recirculation region detaches, and, as fluid bleeding is increased, moves away from the body rear. When bleeding exceeds a critical value, the region suddenly breaks down and disappears \cite{leal1969effect,castro1971wake,lee1999laboratory,yaragal2002two,cummins2017effect,basnet2017structure,sevilla2004vortex}.

This behaviour has profound effects on the bluff-body drag and stability of the flow \citep{ledda2018suppression,sevilla2004vortex}. It thus comes as no surprise that several natural organisms have evolved to possess porous membranes \citep{cummins2018separated} and wings \citep{Sunada2737}, which control the recirculation region and contribute to a more stable and efficient flight. In engineering, base bleed is taken into account when designing efficient airfoils \citep{bearman1967effect,leal1969effect}, dynamic mixers \citep{steiros2017effect}, or when modelling wind turbines and wind farms \citep{Hansen2008,ayati2019double}, to name a few applications. 

 
Despite the above, the reasons behind the emergence, evolution and disappearance of the detached recirculation region are not clear. The recirculation can be expected to be connected to the intense vorticity of the shear layers which induce backflow velocities in the wake \cite{ledda2018suppression}. However, the vorticity magnitude alone is insufficient to act as a causal explanation for the state the flow system settles on. As a matter of fact, several potential flow bluff-body models can be found in the literature, the classical model of \citet{kirchhoff1869theorie} being one example, which predict large values of shear layer vorticity, but do not include a recirculation bubble in their predictions \ks{(for extensions of the Kirchhoff model with a vortical wake see \cite{chernyshenko1988asymptotic,chernyshenko1993high})}. Therefore, apart from the magnitude of vorticity, an explanation of the recirculation region should include at least some information on the distribution of this vorticity, which in turn can be thought to be a function of the diffusion process of the flow (turbulent or molecular) and various constraints (e.g. momentum and mass conservation). This fact was pointed out in the early investigations of \citet{bearman1967effect} and \citet{leal1969effect} who postulated that the recirculation region is a product of the entrainment process of the free shear layers.

In this paper, we use a combination of potential flow modelling, experiments and numerical simulations, and attempt a qualitative explanation on why the recirculation emerges, migrates away from the porous body and disappears. To perform that, we consider the simplest wake with base bleed, that of an infinite-aspect-ratio porous plate, and create a theoretical model which, despite its simplicity, is able to capture, qualitatively, certain basic dimensions and trends of the bubble. The analysis \ks{is strictly valid only for the special case of steady wakes, but our arguments, together with the offered insights, can be thought to also be relevant to the more complex and/or unsteady} configurations that appear in nature and engineering.

The structure of the article is as follows. Section II introduces the phenomenology of the problem, and discusses some major assumptions of the ensuing analysis. Section III presents the novel analytical model for the porous plate recirculation region, and compares its predictions with experimental and numerical data. Finally, section IV summarizes the results, and discusses the extensions of this study.

\begin{figure*}
\centerline{
	\includegraphics[width=\columnwidth]{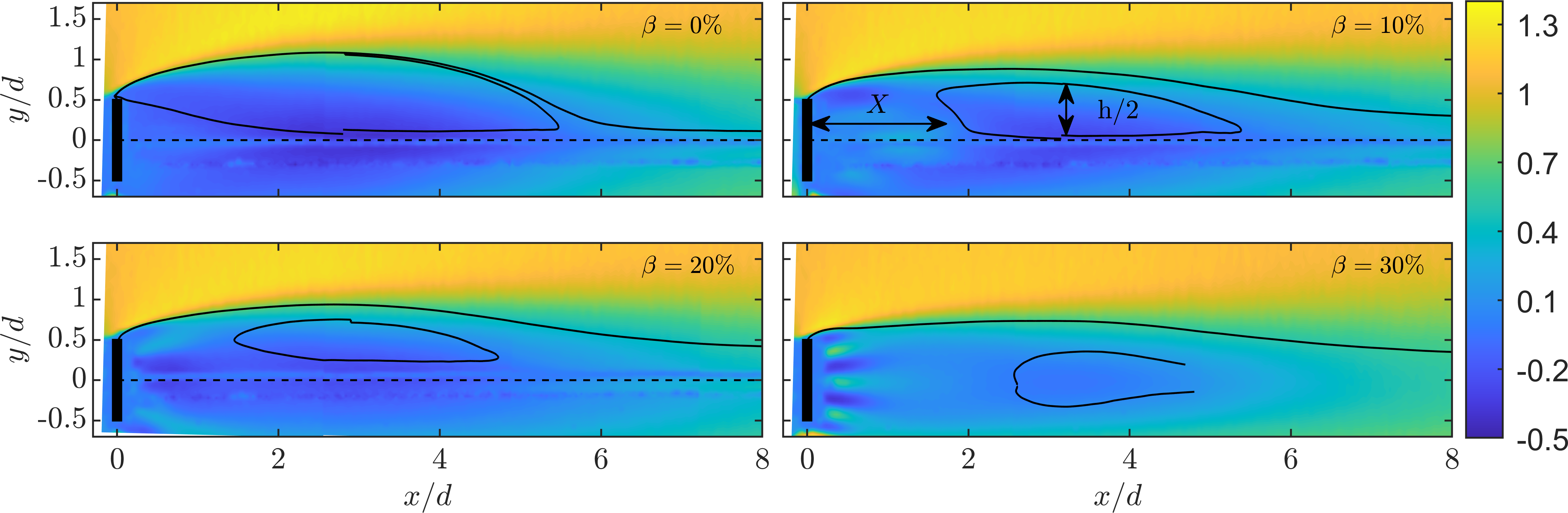}
}
\caption{Steady wake bubble boundary and slipstream of plates of variable porosity, $\beta$, as measured using PIV at $Re = 6\times10^3$. The \kss{colorbar} corresponds to the time-averaged non-dimensional streamwise velocity. $X$ is the bubble detachment length and $h/2$ its maximum half-height. \kss{$\beta$ is the open area ratio of the plate}. The dashed line signifies the presence of a splitter plate, which was used to stabilize the wake in the cases $\beta=0$, $0.1$ and $0.2$. For $\beta=0.3$ vortex shedding is already suppressed by the large base bleeding \citep{castro1971wake}, rendering the splitter plate superfluous.}
\label{fig:PIV1}
\end{figure*}

\section{Background and preliminary considerations}
\label{sec:Preliminary}
Figure \ref{fig:PIV1} shows the recirculation bubble of four plates of different porosities for the fully turbulent regime ($Re = 6\times10^3$ based on the plate height $d$ and free-stream velocity $U_\infty$), measured using Particle Image Velocimetry (PIV) (the details of the experimental set-up are provided in appendix \ref{app:exp}). The depicted flow fields correspond to ``steady" wakes, where vortex shedding is suppressed, either with the use of a splitter plate (cases $\beta = 0, \, 0.1, \, 0.2$, where $\beta$ is the open area ratio defined as the porous plate area over the gross plate area), or from the large base bleeding (case $\beta=0.3$), which acts as an effective splitter plate \citep{castro1971wake}. We note that for the case $\beta=0.3$ the bubble streamlines did not reconverge to a stagnation point (possibly due to 3D effects), and for this reason we chose to depict the bubble as open. We also note that regardless \ks{of} the stabilization technique (splitter plate or wake bleeding), the shear layers will eventually become unstable if allowed to interact, and will generate vortex packets. However, these will be much weaker and will be formed far downstream of the body, compared to the the case where no stabilization is used, so that their effect in the near wake can be thought to be negligible (see discussion in \citet{Steiros2020}). 

Similar to past investigations \citep{castro1971wake}, it is observed that as base bleeding increases, the recirculation bubble detaches from the plate and migrates away from the body, shrinking in size, until it abruptly disappears at a critical open area ratio between $\beta =0.3$ and $\beta =0.4$ (no recirculation bubble was observed at $\beta=0.4$). An analogous behaviour is observed in the laminar regime \citep{leal1969effect}, with the difference that the detachment length $X$ increases linearly with $Re$, whereas in turbulent conditions the bubble dimensions are generally thought to be $Re$-independent. On the other hand there are indications that the bubble height is $Re$-independent even in the laminar regime \citep{leal1969effect}.

Our aim is to provide an explanation \ks{for} the above observations, and create a model which can capture, qualitatively, the dimensions $X$ and $h$, and the porosity range where the bubble exists. Our starting point is the porous plate model of \citet{steiros2018drag}, which is based on potential flow theory, ``corrected" using the constraints of mass and momentum conservation. This model does not predict \ks{nor} presupposes a recirculation bubble and shear stresses in the wake. Despite that, it yields accurate predictions of the drag of infinite aspect ratio porous plates, as long as vortex shedding is absent or suppressed. This model also yields accurate predictions for the wake pressure, $p_w$, at least for the case where $\beta =0$. Moreover, the external flow velocities (i.e. outside the wake) have been shown \citep{Graham1976} to be described reasonably well by the potential flow description that \citet{steiros2018drag} use, which is identical to the one presented in \citet{taylor1944air} when considering the outer flow. 

\begin{figure}
\centerline{
		\includegraphics[width=0.45\columnwidth]{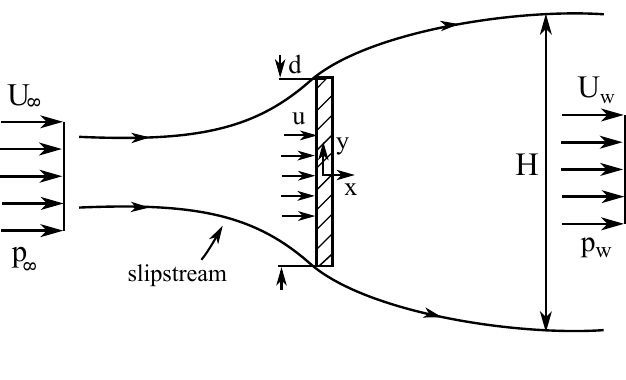} 
}
\caption{Potential flow representation of the steady wake by \citet{steiros2018drag}. $U_\infty$ and $p_\infty$ denote the free-stream velocity and ambient pressure respectively. $U_w$ and $p_w$ denote the wake velocity and pressure after the initial wake expansion. $H$ is the wake height after the initial expansion, and $d$ is the plate height. $u$ is the velocity immediately upstream/downstream of the plate.}
\label{fig:Model1}
\end{figure}

Given the above, we \ks{make} the assumption that inclusion of the recirculation bubble and shear stresses in the analysis will only alter the wake velocities, and will not affect the plate drag, wake pressure, and wake geometry that are predicted by the modified potential flow model of \citet{steiros2018drag}. The prediction of \citet{steiros2018drag} for the wake velocity after the initial wake expansion, $U_w$ (see figure \ref{fig:Model1}), is given below as it is used repeatedly in the analysis that follows:

\begin{equation}
\frac{U_w}{U_\infty} =  \frac{u^*}{2-u^*} \,,
\label{eq:oldMod}
\end{equation}


\noindent where $u^*=u/U_\infty$ is the plate bypass ratio (see figure \ref{fig:Model1} for a definition of the flow quantities). This variable is central to the model of \citet{steiros2018drag}, and is the independent variable of the models developed in the present study. However, in most cases, this is not a readily available quantity. A more straightforward approach would be to express all quantities as a function of the open area ratio $\beta$, and for that an expression linking $\beta$ with $u^*$ is needed. \citet{taylordavies} have provided and validated such an equation when friction losses are negligible. This, together with the drag expression of \citet{steiros2018drag}, yields the following expression (see appendix \ref{app:2} for more details on the derivation)


\begin{equation}
\beta \approx \frac{\sqrt{3(2-u^*)}u^*}{\sqrt{-3u^{*3}+2u^{*2}-4u^*+8}}\,.
\label{eq:bu}
\end{equation}

Note that friction losses can be thought to be negligible when the flow is fully turbulent, or when the plate is vanishingly thin. In that case, the effects of other properties, such as the permeability, can be thought to be secondary (see for instance \cite{Hoerner1952} and \cite{deBray1957}). In the remainder of the text, the above equation will be used to plot the predictions as a function of $\beta$ instead of $u^*$.

\section{Characteristics of the recirculation bubble}
\label{sec:main}

\subsection{Bubble height and disappearance}
\label{sec:width}
\begin{figure}
\centerline{
	\includegraphics[width=0.5\columnwidth]{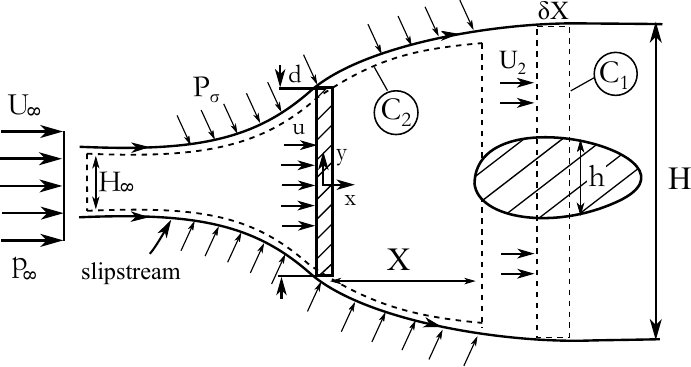} 
}
\caption{Extension of the potential flow representation of the wake of figure \ref{fig:Model1}, by addition of a recirculation region. The dashed regions C$_1$ and C$_2$ denote control volumes. }
\label{fig:Model2}
\end{figure}

We assume the flow field shown in figure \ref{fig:Model2}, where a recirculation bubble is superimposed on the potential flow model depicted in figure \ref{fig:Model1}. The proposed flow model can be realistic only if it is consistent with conservation of mass and momentum. We thus require that these constraints be fulfilled in control volume C$_1$, which is positioned at a location where the bubble is sufficiently parallel to the slipstream, so that the characteristic (or $y-$averaged) flow velocity $U_2$ between the bubble and the slipstream can be assumed to be approximately constant with the streamwise distance $x$.
\kss{Figure \ref{fig:PIV1} shows that this condition is locally satisfied around the location where the bubble assumes its maximum height.}

For sufficiently high values of $Re$, the shear stress on the slipstream or the bubble streamlines is mainly the Reynolds stress $\rho \overline{u' v'}$. Following \citet{roshko1993perspectives}, this can be modelled as $C_t \rho (U_a - U_b)^2$, where $U_a$ and $U_b$ are the characteristic velocities at the two sides of the shear layer, and $C_t$ is a coefficient of proportionality which is assumed constant at a first approximation. Application of momentum conservation on control volume C$_1$ and on the bubble yields

\begin{align}
\begin{split}
2C_t \rho(U_\infty - U_2)^2 \delta X & = \delta P H \, ,\\
2C_t \rho U_2^2 \delta X & =  \delta P h \,  ,
\end{split}
\label{eq:width1}
\end{align}

\noindent where \ks{$\delta X$ is the length of control volume C$_1$, \kss{$\delta P$ is the pressure increase inside the control volume $C_1$} and the factor two is due to the contribution of both (upper and lower) shear layers. In the above it was assumed that the velocity outside the wake is $U_\infty$, and zero inside the bubble, the latter following \citet{roshko1993perspectives}. These velocity values are not strictly correct (see figure \ref{fig:PIV1}) and should only be considered as characteristic of the corresponding mixing layers. Different velocity values to the ones assumed would alter, slightly, the predicted height. The predicted trends and general arguments would, however, remain unaltered.}

If the bubble were not present, the streamwise velocity at C$_1$ would be predicted from the potential flow model of \citet{steiros2018drag} to be $U_w$, i.e. the velocity of the expanded wake, given by equation \eqref{eq:oldMod}. With the addition of the bubble in the analysis, the cross-sectional area that the fluid may pass \ks{through} is reduced, and the new velocity, $U_2$, can be calculated using mass conservation as


\begin{equation}
\rho U_2 (H-h) = \rho U_w H \,.
\label{eq:U2}
\end{equation}

Combination of equations \eqref{eq:oldMod}, \eqref{eq:width1} and \eqref{eq:U2} yields

\begin{equation}
\frac{h}{H} = \left( \frac{u^*}{\left(1 - \frac{h}{H}\right)(2-u^*) - u^*} \right)^2 \,,
\end{equation}

\noindent which links the non-dimensional maximum bubble height $h/H$ \kss{(i.e. at the location where the bubble is parallel to the slipstream)} with the bypass ratio $u^*$ or, equivalently, with the open area ratio $\beta$. The only physically relevant solution to the above equation is 

\begin{equation}
\frac{h}{H} = \frac{3u^* - 2 - \sqrt{(u^*-2)(5u^*-2)}}{2(u^*-2)} \,.
\label{eq:h2}
\end{equation}

The predictions of equation \eqref{eq:h2} are plotted in figure \ref{fig:Res1}(a), together with values measured from the PIV experiments and Large-Eddy Simulations (see appendix \ref{app:method} for details). The observations are in qualitative agreement with the theoretical predictions. When \ks{ $0.4 < u^* \leq 1$}, equation \eqref{eq:h2} does not yield real values: the model-bubble is incompatible with the constraints imposed by momentum and mass conservation. More specifically, the system \eqref{eq:width1} shows that the external shear generates a pressure gradient in the wake, and thus constrains the bubble shear, as the latter must also produce the same pressure gradient. The above can only occur when the bubble has a specific height as this affects on one hand the shear (by changing the velocity between the slipstream and the bubble $U_2$) and on the other the pressure force on the bubble. Above a critical porosity value no bubble is compatible with the above constraints, and it must therefore disappear. Our simplified model predicts this disappearance to occur at $\beta\approx 0.34$ (which is equivalent to $u^*=0.4$, according to expression \eqref{eq:bu}). In agreement with that prediction, our measurements indicate that the ``critical" porosity $\beta$ lies somewhere between 0.3 and 0.4, while our simulations predict a critical porosity of $\beta=0.35$, where the recirculation bubble appears intermittently in time, and is absent when considering the time-averaged flow field. 

\begin{figure*}
\centerline{
	\begin{tabular}{ll}
		(a) & $\qquad$ (b) \\
		\includegraphics[width=.48\columnwidth]{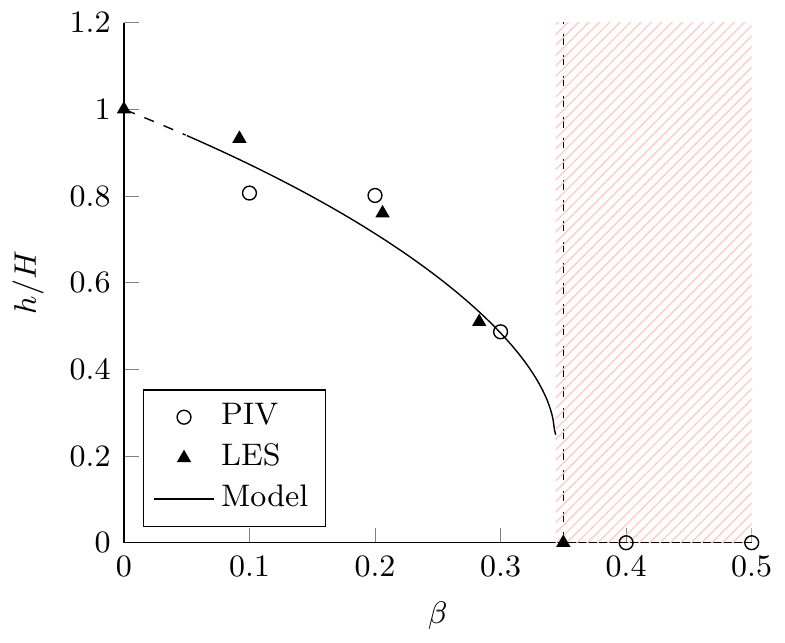} &
		\includegraphics[width=.48\columnwidth]{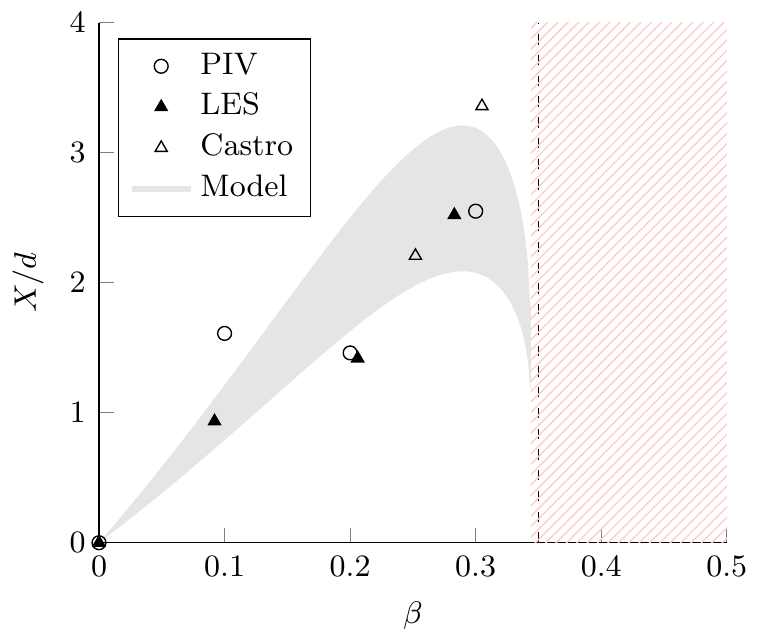}
	\end{tabular}
}
\caption{Normalized (a) bubble height and (b) detachment length, as a function of $\beta$. \kss{In the dashed part of the prediction, where $\beta \to 0$, the theoretical analysis is not strictly valid}. The hatched areas represent the region where the mass and momentum conservation laws cannot be fulfilled. The dashed vertical line indicates the porosity where the recirculation bubble disappears, according to the numerical simulations. Where available, results from the experiments of \citet{castro1971wake} are also included. The ratio $h/H$ was measured at the location of maximum bubble height, while the distance $X$ is the minimum distance between the bubble and the plate.}
\label{fig:Res1}
\end{figure*}

It must be mentioned that when $\beta \to 0$ (and $h \to H$), there is not enough separation between the bubble boundary and the slipstream to consider them as two separate shear layers (see figure \ref{fig:Model2}). In that case, the system of equations \eqref{eq:width1}, and thus the above analysis, is not valid. However, equation \eqref{eq:h2} yields the correct prediction of $h/H=1$ for $\beta=0$, while we expect the region of invalidity to be small, as for $\beta=0.1$ the separation of the bubble and slipstream is already of significant extent (see figure \ref{fig:PIV1}). We thus use equation \eqref{eq:h2} even when $\beta \to 0$, and we expect that the deviation from the ``correct" prediction is small. To emphasize the above, the part of the prediction where we expect it to not be strictly valid has been marked with a dashed line in figure \ref{fig:Res1}(a). 

\subsection{Bubble detachment and emergence}
\label{sec:detach}

We now consider control volume C$_2$ (see figure \ref{fig:Model2}), which is bounded by the slipstream, and extends from far upstream of the plate to the tip of the bubble, i.e. at distance $X$ from the plate. As argued in section \ref{sec:Preliminary}, we assume that the pressure at this location is equal to $p_w$, i.e. the wake pressure as predicted by \citet{steiros2018drag}. If the bubble and shear stresses were absent from the analysis, the problem would be reduced to the one of \citet{steiros2018drag}, and in that case, a momentum balance in C$_2$ would yield
 
\begin{equation}
D = \rho u d (U_\infty-U_w) + p_\infty H_\infty - p_w H + P_\sigma  \,,
\label{eq:X1a}
\end{equation}

\noindent where $D$ is the drag and $P_\sigma$ is the pressure force on the slipstream. \citet{steiros2018drag} employed a slightly different control volume, but under identical assumptions, and their results are thus equivalent. 

We now increase the complexity of the analysis by adding a bubble and shear stress on the slipstream. As argued in section \ref{sec:Preliminary}, it is assumed that only the wake velocities are altered from the added complexity. Thus, a momentum balance in C$_2$ now yields

\begin{equation}
D = \rho u d (U_\infty-U_2) + p_\infty H_\infty - p_wH + P_\sigma + F_s  \,.
\label{eq:X1b}
\end{equation}

For high $Re$, the forcing on the slipstream $F_s$ is mainly the Reynolds stress. Similar to the discussion in section \ref{sec:width}, this is modelled as being proportional to the square of the difference of the characteristic velocities across the slipstream. An approximation is to consider the outside velocity to be $U_\infty$, and the inside one to be $u$. Thus, the force due to shear stress becomes

\begin{equation}
F_s \approx 2\rho C_t' (U_\infty - u)^2 X  \,,
\label{eq:X1c}
\end{equation}

\noindent where the factor 2 is due to the two (upper and lower) shear layers, and $C_t'$ an empirical parameter (provided from experiments or numerical simulations). Combining equations \eqref{eq:X1a}, \eqref{eq:X1b} and \eqref{eq:X1c} we obtain

\begin{equation}
2 \rho C_t' (U_\infty - u)^2 X \approx \rho ud(U_2 - U_w)  \,.
\label{eq:X1d}
\end{equation}

We interpret the above equation by following \ks{an argument analogous to that of} \citet{bearman1967effect} and \citet{leal1969effect}: immediately downstream of the plate, the shear on the slipstream will cause an acceleration of the wake fluid which is closest to the slipstream and, in order for mass to be conserved, the fluid at the core of the wake will have to be continually decelerated. When it reaches zero velocity, a recirculation bubble will form. At this exact location, the momentum flux will be approximately $\rho ud U_2$, i.e. increased by $\rho ud(U_2 - U_w)$ compared with the case where a bubble is not included in the flow description. Our simplified analysis assumes that this increase in momentum flux is exactly provided by the shear stress on the slipstream.


Combination of equations \eqref{eq:oldMod}, \eqref{eq:U2}, \eqref{eq:h2} and \eqref{eq:X1d}, leads to the following expression for the normalized detachment length $X/d$

\begin{equation}
\frac{X}{d} = \frac{1}{2C_t'} \frac{{u^*}^2}{(1-u^*)^2(2-u^*)} \frac{3u^*-2-\sqrt{(u^*-2)(5u^*-2))}}{\sqrt{(u^*-2)(5u^*-2))}-u^*-2}  \,,
\label{eq:X1e}
\end{equation}

\noindent where $C_t'$ must be determined from the shear stress on the part of the slipstream which extends from the plate to the start of the bubble. As described in appendix \ref{app:method}, by integrating the shear stresses on the slipstream up until the bubble, and then averaging across all tested cases, we obtain a value of  $C_t' = 0.0257$ for our laboratory, and $C_t' =0.0395$ for our numerical experiments. These values are somewhat different, but we note that our region of integration lies very close to the plate, where the flow is extremely inhomogeneous and sensitive to boundary conditions, which are slightly different in the simulations and experiments. At larger downstream distances the initial conditions (e.g. shape of holes) are expected to not affect the statistics, as shown in the comparison between experimental and numerical results (see appendix \ref{app:num}).

In figure \ref{fig:Res1}(b) we plot equation \eqref{eq:X1e} using the above two values of $C_t'$ to form an envelope. We observe that the predictions are in qualitative agreement with the computed/measured values of $X/d$, suggesting that the arguments of the model are not far from what occurs in reality.

\subsection{Laminar regime}
\label{sec:laminar}

The analysis presented in the above sections concerns the fully turbulent regime, where the bubble dimensions are expected to be independent of the Reynolds number, in accordance with equations \eqref{eq:h2} and \eqref{eq:X1e}. In the laminar regime on the other hand, \citet{leal1969effect} observed that while the normalized height of the centre of the bubble remains $Re$-independent, the detachment length $X$ increases linearly with $Re$. We will attemt to explain these observations using the proposed framework.

We consider a steady laminar wake and assume that the present analysis is capable of describing the flow field. At first this seems counter-intuitive, given that one would expect that an analysis based on potential flow theory is likely to be valid only for very high Reynolds numbers. Nevertheless, the arguments and results of \citet{acrivos1965steady} and \citet{leal1969effect} indicate that an ``infinite" $Re$ analysis is appropriate for bluff bodies whose Reynolds numbers are as low as 25.

The analysis presented in sections \ref{sec:width} and \ref{sec:detach} may thus be repeated, with the only difference being that the stresses on the slipstream and bubble are now predominantly viscous rather than turbulent. These can be approximately modelled using a standard boundary layer analysis with $l$ as the characteristic length

\begin{equation}
F_s = C_v \mu \frac{U_a - U_b}{l}\sqrt{Re_l}\,,
\end{equation}

\noindent where $\mu$ is the dynamic viscosity of the fluid, $(U_a-U_b)$ is the velocity difference across the shear layer and $C_v$ is a coefficient of proportionality. A momentum budget similar to the one presented in section \ref{sec:width} now yields the following system of equations

\begin{align}
\begin{split}
2C_v \mu \frac{U_\infty - U_2}{l_1Re_{l_1}^{-1/2}} \delta X & = \delta P H \, ,\\
2C_v \mu \frac{U_2}{l_2Re_{l_2}^{-1/2}} \delta X & =  \delta P h \,  .
\end{split}
\label{eq:width2}
\end{align}

\noindent where $l_1$ and $l_2$ are the boundary-layer characteristic lengths for the slipstream and bubble, respectively.

\begin{figure*}
\centerline{
	\begin{tabular}{ll}
		(a) & $\qquad$ (b) \\
		\includegraphics[width=.48\columnwidth]{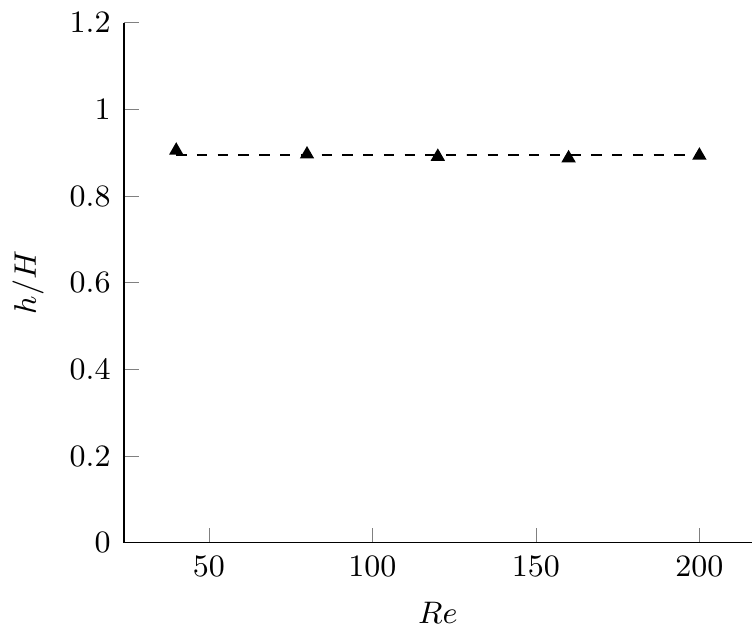} &
		\includegraphics[width=.48\columnwidth]{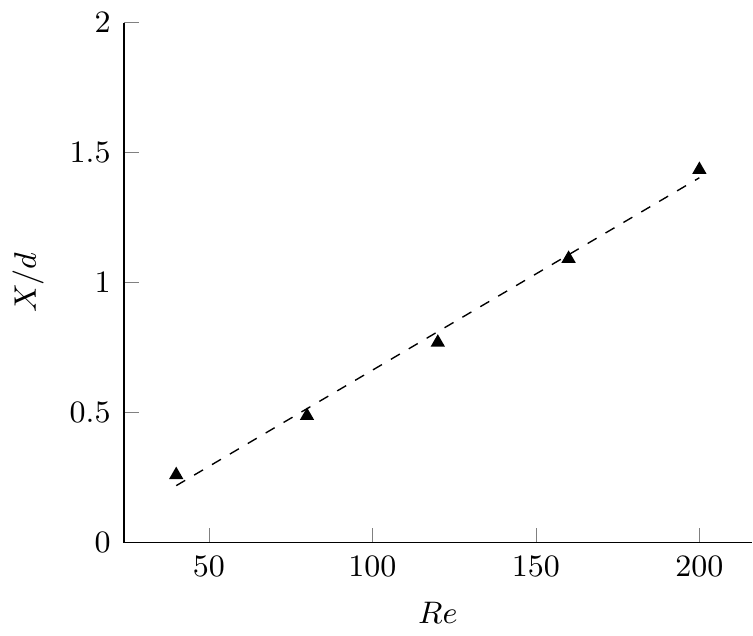}
	\end{tabular}
}
\caption{Normalized (a) maximum bubble height and (b) detachment length, as a function of $Re$ in the laminar regime, for a thin plate where $\beta=20.6\%$. The ratio $h/H$ was measured at the location of maximum bubble height, and the distance $X$ as the minimum distance between the bubble and the plate. Symbols: DNS. Line: linear fit.}
\label{fig:Res2}
\end{figure*}

Calculation of the normalized bubble height $h/H$ requires taking the ratio of the above equations, and thus any dependency on viscosity cancels out, as per the observations of \citet{leal1969effect}. 



With respect to the detachment length, the laminar equivalent of equation \eqref{eq:X1d} is 
\begin{equation}
2  C_v \nu (U_\infty - u) \sqrt{\frac{U_\infty X}{\nu}} = ud(U_2 - U_w)  \,,
\end{equation}

\noindent where $\nu=\mu/\rho$ and the characteristic length is taken equal to the detachment distance $X$. This can be rearranged to yield


\begin{equation}
\frac{X}{d} = Re f(u^*) \,,
\label{eq:detLam}
\end{equation}

\noindent which predicts a linear increase of the detachment length with $Re$ in the laminar regime, as per the observations of \citet{leal1969effect}. 

In figure \ref{fig:Res2} we examine the validity of the predicted trends by comparing them with data from a series of Direct Numerical Simulations (DNS) of a plate of $\beta=20.6\%$, for $40 \leq Re \leq 200$ (see appendix \ref{app:method} for numerical details). The simulated plate was very thin ($\approx 1.77\%$ of the plate height $d$), so as to minimize viscous effects due to the friction inside the holes, which can be significant in the laminar regime, and may thus induce effects which are not taken into account in the analysis. In accordance with our theoretical predictions, the normalized bubble height is independent of the Reynolds number, while the detachment length grows linearly with $Re$. 

\section{Concluding Discussion}
\label{sec:Conc}

\begin{figure*}
\centerline{
	\includegraphics[width=\columnwidth]{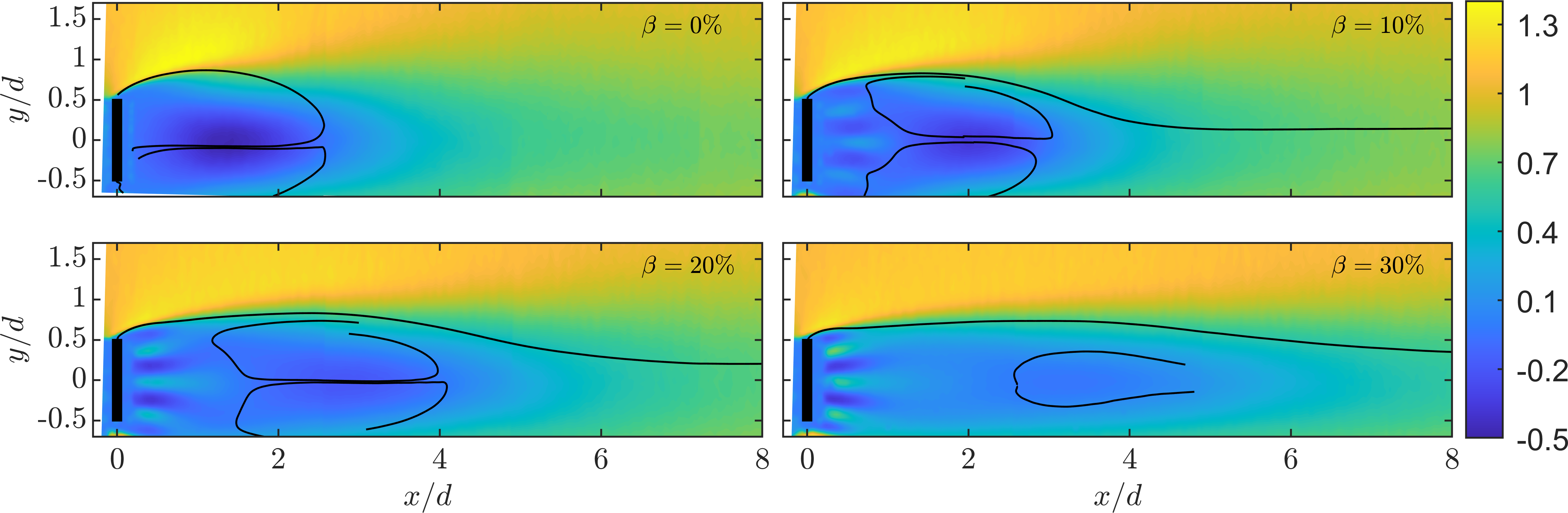}
}
\caption{Time-averaged recirculation bubble for different values of porosity $\beta$, as measured using PIV (see appendix \ref{app:exp}), at $Re = 6 \times 10^3$. The colours correspond to the time-averaged non-dimensional streamwise velocity. Vortex shedding makes the wake unsteady for $\beta = 0$, $0.1$ and $0.2$, but is suppressed for $\beta = 0.3$ due to the large bleeding \citep{castro1971wake}. The bubble evolution is qualitatively similar to the case where the wake is stabilized using a splitter plate (figure \ref{fig:PIV1}).}
\label{fig:PIV2}
\end{figure*}

This work introduces a model for the prediction of the height and detachment length of the recirculation bubble that appears in the steady wake of porous plates when their open area ratio is relatively small. The model extends the work of \citet{steiros2018drag} for drag and flow field of porous plates, by superimposing a recirculation bubble on their predicted flow field. \kss{The main assumptions are that (i) the flow field is steady and two-dimensional and can be well-represented using potential flow physics; (ii) the addition of stresses (turbulent or viscous) on the wake slipstream modify the potential flow description only by altering the velocities inside the wake, leaving the pressure, plate forces and outer flow unaffected; (iii) the characteristic velocity inside the bubble is zero; (iv) the bubble boundary is parallel to the slipstream at the position of maximum bubble height; and that (v) the slipstream stresses can be modelled as canonical free shear layers (turbulent stresses) or laminar boundary layers (viscous stresses).}

On the basis of the above theoretical analysis, we attempt interpretations on why the bubble (i) emerges, (ii) moves downstream with increasing base bleed and (iii) disappears at a critical value of porosity. Moreover, by applying the developed framework in both laminar and turbulent regimes, we predict that the normalized height of the bubble is always $Re$-independent, while the detachment region grows linearly with $Re$ in the laminar regime, but is $Re$-independent in the turbulent regime, in accordance with past and present observations.

The above analysis concerns the steady wake of infinite aspect ratio porous plates. Nevertheless, the physical arguments that are presented in this study, along with the various explanations and interpretations, are general, and can be thought to be applicable to the detached recirculation regions that characterize many other wakes with base bleed that appear in nature and engineering. A pertinent example is those wakes in which bleeding is generated by actively injecting fluid in the wake, rather than passively perforating the body. Such apparatuses are relevant, for instance, to bluff airfoil flow control \citep{bearman1967effect,leal1969effect}. In such cases, a recirculation bubble again appears, which shares all the characteristics and trends of the recirculation bubble of porous plates, albeit possessing different characteristic dimensions. Another example is the wake of various flowers and seeds. Despite their complexity, these organisms generate wakes and detached recirculation regions which are remarkably similar to those of porous plates \citep{cummins2018separated}.

When the wake becomes unsteady due to the emergence of vortex shedding, the time-averaged flow picture exhibits a recirculation bubble, whose trends \ks{and} physics \kss{show some similarity to the ones of the steady case (see figure \ref{fig:PIV2}). It is conceivable, then, that the arguments of the present study could still be applicable} in a qualitative sense, even though the actual wake dimensions cannot be accurately described by the theory presented here, which concerns steady wakes.

\begin{acknowledgments}
Part of this work was conceived while the first author was a post-doctoral associate at Princeton University. We are grateful to Prof. Marcus Hultmark  for helpful comments and discussions, and to Prof. Lex Smits for providing access to the water channel where the experiments were conducted. 
\end{acknowledgments}

\appendix
\section{Details of experimental and numerical methods}
\label{app:method}
\subsection{Experimental apparatus and data processing}
\label{app:exp}

\begin{figure}
\centerline{
\begin{tabular}{ll}
    \multicolumn{2}{l}{(a)} \\
    \multicolumn{2}{c}{\includegraphics[width=.6\columnwidth]{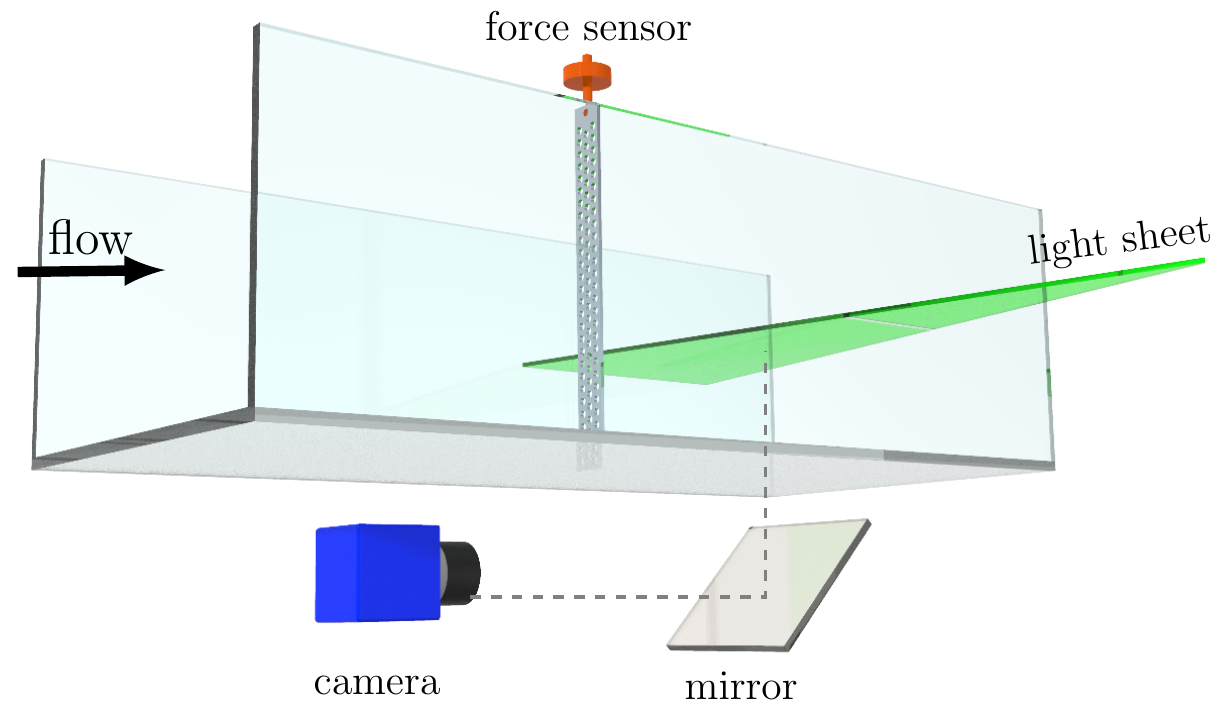}} \\
	(b) & $\qquad$ (c) \\
\includegraphics[height=0.2\columnwidth]{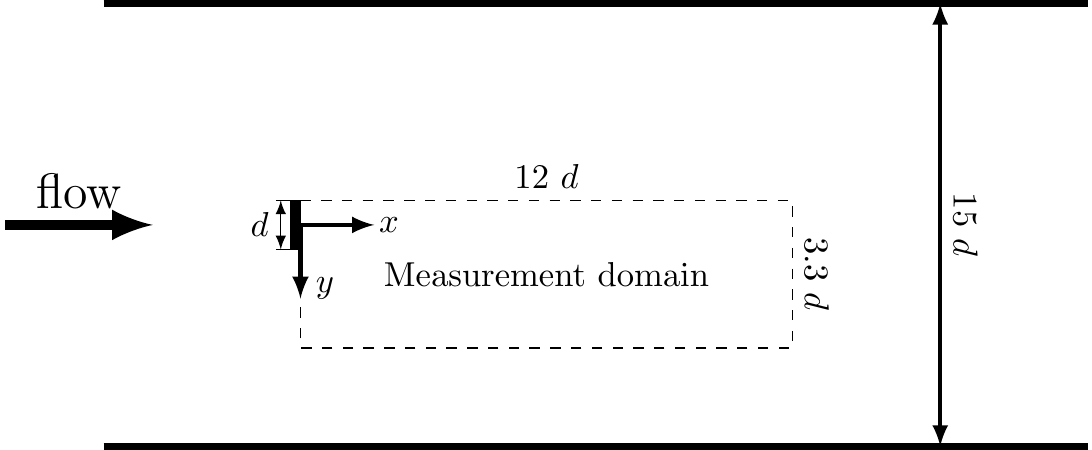} & $\qquad$
\includegraphics[height=0.225\columnwidth, trim=0cm 1cm 0cm 0cm]{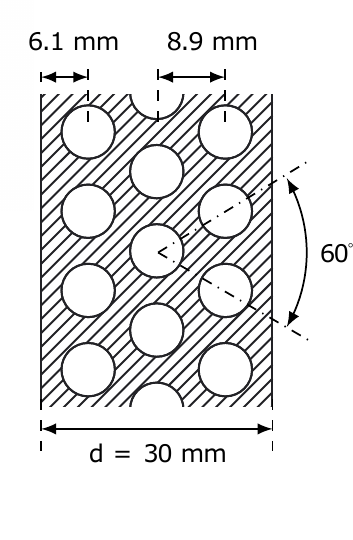}
\end{tabular}
}
\caption{\kss{(a) Side and (b) top view of the experimental configuration, and (c) plate design. All tested plates have the same hole-pattern, but different hole diameter}.}
\label{fig:exp_setup}
\end{figure}

The validation experiments were conducted in a free-surface recirculating water tunnel. The tunnel had a 0.46 m wide, 0.3 m deep and 2.44 m long acrylic test section, and operated at a constant free-stream velocity of 0.2 ms$^{-1}$. The experimental set-up is shown in figure \ref{fig:exp_setup}.

Porous plates with porosities $\beta=$ 0\% (solid), 10\%, 20\%, \ldots, 60\% porosity were tested. The plates were 29 cm long, 3 cm wide and 0.3 cm thick, \kss{were made from aluminium} and were perforated with circular holes arranged in the way depicted in figure \ref{fig:exp_setup}. \kss{The hole diameter was varied among the values 3.6, 5.1, 6.2, 7.2, 8.1 and 8.8 mm in order for different porosities to be achieved. The channel blockage was $6.5 \%$ for the solid plate and lower for the porous plates. The Reynolds number, based on the plate height, $Re = U_\infty d/\nu$, was 6,000 for all plates.} In the experiments, the plates were supported from the top and were positioned near the center of the test section. The plates touched the floor of the channel at one end, while the other end extended above the water free surface, in order to suppress three-dimensional effects arising from the plate edges. For porosities 0\%, 10\% and 20\%, a 2 mm thick splitter plate was positioned at the wake of the plate, spanning approximately $10$ porous plate heights. This was done to stabilize the wake by suppressing the vortex shedding which otherwise appears at these porosities \citep{castro1971wake}. \kss{We note that a splitter plate generates additional effects in the flow field, as for instance skin friction, which are not included in the model. However, the PIV data show relative agreement with both the theoretical model and the LES data (symmentry, no splitter plate, see section \ref{app:num}), suggesting that these effects can be largely neglected. Utilization of a splitter plate as a validation tool for steady-wake models has been repeatedly employed in past works (see for instance \cite{roshko1993perspectives,acrivos1965steady})}.

Planar PIV was performed to study the evolution of the recirculation bubble. A dual-cavity Nd:YAG laser (Litron Nano PIV, 50 mJ pulses) was used to form a light sheet at half depth of the tunnel. A PIV camera (LaVision Imager sCMOS) was situated underneath the tunnel to image the wake through the reflection of a mirror. A 2$\times$ teleconverter (Vivitar) was attached to the camera yielding a magnification of 0.16. In the cross-stream direction, the field of view covered the full height of the plate plus another 2.3 plate heights on a single side of the plate. In the streamwise direction, velocity data were acquired at five stations that spanned over 12 plate heights downstream of the plate. Different downstream stations were accessed by traversing the plate upstream while keeping the camera and mirror fixed, since the free stream in the middle portion of the test section was homogeneous. A small overlap region between the measurement planes allowed their subsequent stitching, by finding the location where the velocity profiles from two different planes collapsed in an optimal manner.

At each station, 1,000 image pairs were recorded and analyzed to calculate flow statistics. PIV interrogation was performed in DaVis 8.3 using a multi-grid, image deformation strategy \citep{scarano2001iterative}, with a final interrogation spot size of 2 mm $\times$ 2 mm and 50\% window overlap. Velocity vectors were validated with the universal outlier detection method \citep{westerweel2005universal}. 

The bubble boundary was found by locating the \kss{streamlines that reversed their direction in the averaged flow field}, and then choosing the one that created the largest bubble. It was often the case that two streamlines were stitched together to form the bubble boundary: one for the front change of direction (bubble tip), and one for the rear. The slipstream was found by locating the streamline that passed from the plate edge. The empirical coefficient $C_t'$ in the discussion of section \ref{sec:detach} was found by calculating the normalized turbulent stress ${\overline{u'v'}}/{(U_\infty - u)^2}$ on the slipstream, from a location 0.15$d$ downstream of the plate, to the location where the bubble emerged, and then taking its average value ($u^*$ was calculated using equation \eqref{eq:bu}). The downstream distance 0.15$d$ was chosen as a compromise between starting the averaging process as close as possible to the plate, and far enough, in order to minimize the noise due to reflections and shadows. For the cases where a steady detached bubble emerged, (i.e. $\beta=0.1$ and 0.2 with splitter plate and $\beta=0.3$ without splitter plate), $C_t'$ was calculated equal to 0.0257 $\pm$ 0.002.

\subsection{Numerical simulations}
\label{app:num}
Simulations were performed using the general-purpose finite-volume toolbox OpenFOAM. For the turbulent flow cases, where the Reynolds number based on the plate height and free-stream velocity was $Re=6240$, Large-Eddy Simulations (LES) using the dynamic one-equation subgrid-scale model of \citet{kim1995new} were performed. For the simulations in the laminar regime (section \ref{sec:laminar}), the sub-grid scale model was dropped, and Direct Numerical Simulations (DNS) were conducted. The convective terms were discretized with a second-order central differencing scheme. A second-order backward scheme was used for time discretization. 

The porosity of the plates was controlled by changing the size of the homogeneously distributed square-shaped holes. \kss{The arrangement of the holes mimicked that of the experiments (shown in Fig. \ref{fig:exp_setup}), but their shape was different to facilitate the grid generation process.} For the simulations in the turbulent regime, the thickness of the plates was similar to those in the experiment, $\approx 10\%$ of the plate height $d$. \ks{Five plates} were considered, with porosities equal to $\beta (\%) =0, 9.2, 20.6, 28.3, 35$. In the laminar flow cases, the plate that was considered was thinner ($\approx 1.77\%$ of the plate height $d$), so as to minimize viscous effects due to the friction inside the plate holes, which can be significant at low Reynolds numbers.

\begin{figure}
	\centering
	\includegraphics[width = .65\columnwidth]{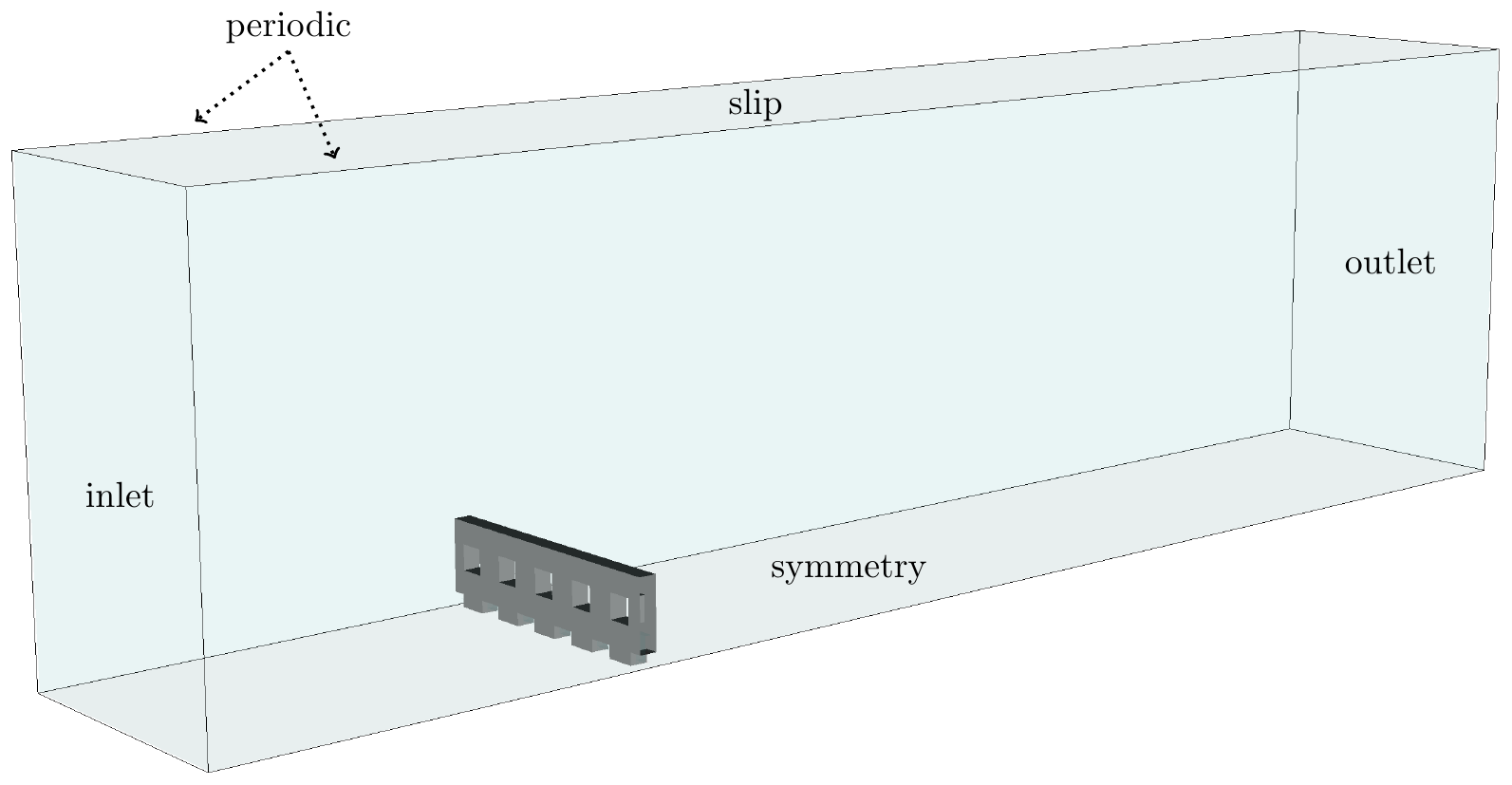}
	\caption{Computational set-up for simulations of the flow past porous plates (domain and plate not drawn to scale).}
	\label{fig:num_setup}
\end{figure}

Communication of the upper and lower layers was inhibited by simulating only half the plate and enforcing symmetry conditions along the centreline. Vortex shedding was therefore suppressed, in a similar way as when a splitter plate is used. The computational domain consisted of a rectangular box extending $27.5 \times 8 \times 1.7$ times the plate height $d$ in the streamwise, normal, and spanwise directions, respectively (see figure \ref{fig:num_setup}). Laminar inflow was imposed at the upstream boundary, whereas periodic boundary conditions were applied in the spanwise direction. The spanwise extent of the domain was taken such that precisely five holes could be accommodated along this direction (see figure \ref{fig:num_setup}). The domain was discretized with $\simeq 3.0\times 10^6$ elements in the turbulent flow cases, and $\simeq 1.8 \times 10^6$ in the laminar ones. In all cases, $50$ cells were used along the span. Data were averaged over a non-dimensional time period corresponding to $t^* = tU_\infty/d = 104$ flow times. For the cases where a steady detached bubble emerged, (i.e. $\beta=0.092$, 0.206 and 0.283), $C_t'$ was calculated as described in appendix \ref{app:exp}, and was equal to 0.0395 $\pm$ 0.0105.

\begin{figure*}
\centerline{
	\begin{tabular}{lll}
		(a) & (b) & (c) \\
		\includegraphics[height=0.25\columnwidth]{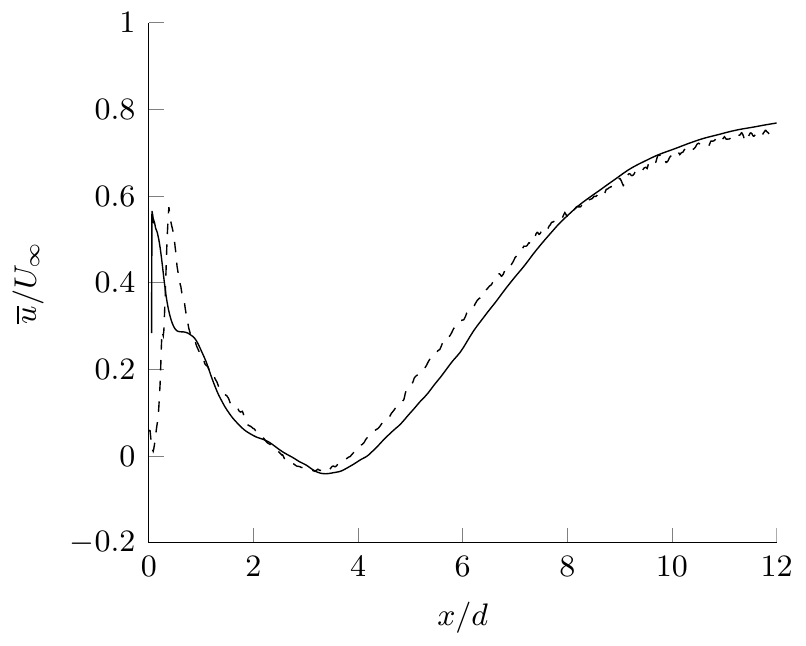} &
		\includegraphics[height=0.25\columnwidth]{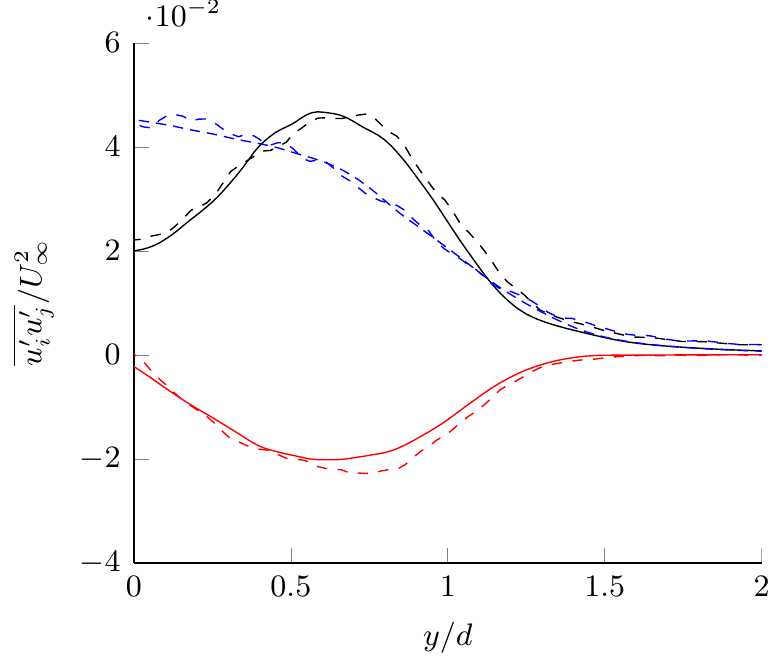} &
		\includegraphics[height=0.25\columnwidth]{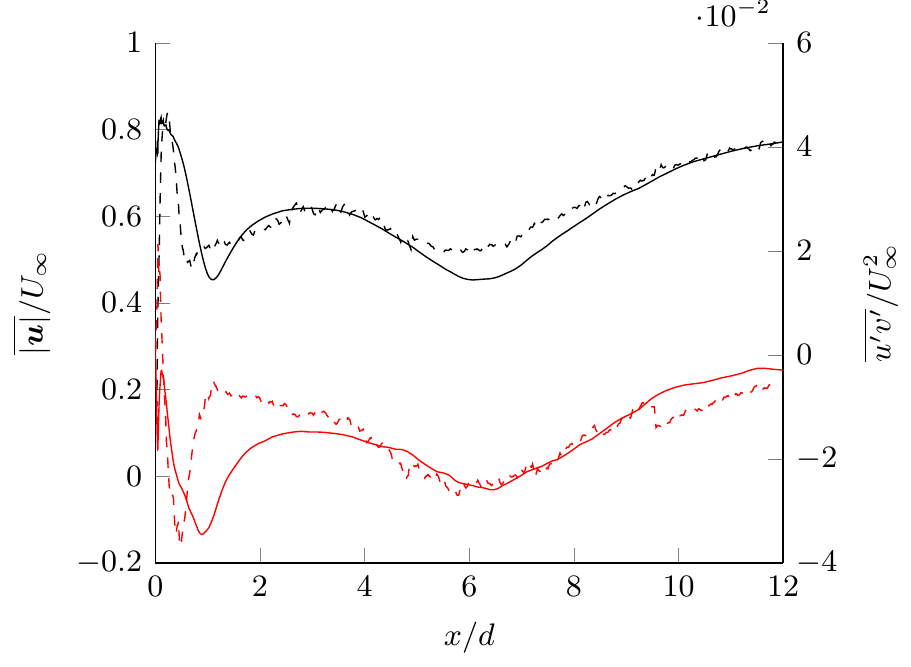}
	\end{tabular}
}
\caption{(a) Streamwise velocity along wake centreline. (b) Reynolds stresses at a cut in the wake at $x/d=5$. Black: $\overline{u'u'}/U_\infty^2$, red: $\overline{u'v'}/U_\infty^2$, blue: $\overline{v'v'}/U_\infty^2$. (c) Black: velocity magnitude and red: Reynolds stress ($\overline{u'v'}/U_\infty^2$) along the slipstream. Time-averaged results. Dashed lines: PIV measurements (from a single plane), solid lines: numerical simulations (averaged over the spanwise extent of the domain).}
\label{fig:valid}
\end{figure*}

Validation of the numerical simulations was performed against the PIV measurements (section \ref{app:exp}), for the case where $\beta = 30\%$ (the plate porosity in the simulations was $\beta = 28.3\%$) and no splitter plate is present in the wake. The symmetry condition was removed, and the domain was doubled in size in the normal direction, so that the entire plate could be accommodated. This change mostly affected the flow downstream of the bubble, whereas the key dimensions of the bubble that are discussed in the main body of the paper were largely unaffected. The results from the laboratory and numerical experiments are compared in figure \ref{fig:valid}; good agreement between the two is observed.  

\section{Derivation of equation \eqref{eq:bu}}
\label{app:2}

\citet{steiros2018drag} provide the following expression for the drag coefficient of an infinite-aspect-ratio plate

\begin{equation}
C_D = \frac{4}{3}\frac{(1-u^*)(2+u^*)}{2-u^*}\,.
\end{equation}

\noindent The above expression provides good agreement with experimental measurements when the wake is steady (i.e. vortex shedding is absent) and the regime fully turbulent.

On the other hand, \citet{taylordavies} propose the following expression, linking $\beta$ to $u^*$, when friction losses are negligible (so that the effect of permeability can be thought secondary)

\begin{equation}
C_D \approx u^{*2}\left(\frac{1}{\beta^2}-1\right)\,.
\end{equation}

The above formula has been repeatedly used in ensuing studies of porous plates \ks{which consider both 3D (finite aspect ratio) and 2D (infinite aspect ratio) geometries} (see for instance \cite{castro1971wake,Graham1976}). Note that the critical \ks{assumptions of the above equation are} that the surplus kinetic energy, due to acceleration of the fluid which enters the plate pores, becomes irreversibly heat, due to the expansion at the end of the pores, \ks{and that the effect of the vena contracta is negligible}. \citet{taylordavies} provide experimental validation for the above formula for a variety of porous plates, in fully turbulent conditions (see also \citet{steiros2018drag} for a reproduction of their validation plot).

Combination of the above two equations yields equation \eqref{eq:bu}.

\bibliography{PorousPlate}

\end{document}